\documentclass[manuscript, screen]{acmart}

\AtBeginDocument{%
  }

\setcopyright{acmlicensed}
\copyrightyear{2025}
\acmYear{2025}
\acmDOI{XXXXXXX.XXXXXXX}
\acmConference[Advancing Post-growth HCI]{Advancing Post-Growth HCI: An ACM CHI 2025 Workshop}{April 25, 2025}{Yokohama, Japan}
\acmISBN{978-1-4503-XXXX-X/2018/06}

\begin{document}

\title{Reflecting on Potentials for Post-Growth Social Media Platform Design}

\author{Joseph S. Schafer}
\email{schaferj@uw.edu}
\orcid{0000-0002-6921-2074}
\affiliation{%
  \institution{University of Washington}
  \city{Seattle}
  \state{Washington}
  \country{USA}
}


\begin{abstract}
  Sudden attention on social media, and how users navigate these contextual shifts, has been a focus of much recent work in social media research. Even when this attention is not harassing, some users experience this sudden growth as overwhelming. In this workshop paper, I outline how growth infuses the design of much of the modern social media platform landscape, and then explore why applying a post-growth lens to platform design could be productive. 
\end{abstract}

\ccsdesc[500]{Human-centered computing~Social media}

\keywords{Attention Dynamics, Slow, Degrowth, Post-growth}

\received{20 February 2025}

\maketitle

\section{Growth, Sudden Attention, and Social Media}
My previous research work has primarily focused on how creators navigate bursts of sudden attention, and how this alters creators' and influencers' perceptions of social media platforms \cite{schafer_post-spotlight_2023, schafer_viral_2023, schafer_i_2025}. This has also been a significant focus of other recent research, across a variety of social media research venues \cite{gurjar_effect_2022, hasan_impact_2022, nahon_going_2013}.  

Users can find these moments of sudden attention overwhelming, due to the rapid nature of their audience growth \cite{schafer_i_2025}, even when these users do not face harassment. Despite this, most mainstream modern social media platforms are primarily oriented around growth, and trying to have their users aim for growth --- via increased followers, views, likes, or time spent on the platform. The rapid proliferation of AI-managed accounts \cite{sato_metas_2025} and AI slop on platforms \cite{diresta_how_2024, yang_anatomy_2024}, consuming excessive amounts of resources while growing these networks with new accounts, is further evidence of the prevalence of pro-growth orientations of social media platforms and their users. 

Uninhibited growth and scaling on social media platforms can cause a variety of harms, from privacy violations \cite{schafer_viral_2023, shi_using_2013}, to the spread of misinformation \cite{wilner_its_2023}, to online harassment \cite{han_hate_2023}. At the same time, social media platforms provide important spaces for social activism \cite{kaviani_bridging_2022,irannejad_bisafar_supporting_2020}, and for finding community, especially for those who are living in marginalized groups in their real-life identities \cite{devito_how_2022, klassen_more_2021}.   

\section{What Post-Growth Could Mean For Social Media}

How social media platforms could be designed to prioritize a post-growth information ecosystem is an open but promising question. In my view, moving design of technology to a post-growth framework also requires a sense of changing temporalities within design. While HCI researchers have recently begun to explore designing for alternative temporalities (e.g. \cite{rapp_introduction_2022, lindley_changing_2013, rapp_how_2022, wiberg_time_2021}), these principles have received less attention than the growth-dominated, accelerating, mainstream platforms. 

Multiple approaches for incorporating these principles into platforms should be used to advance the health of these information ecosystems. Drawing inspiration from activist movements such as the Slow movement \cite{honore_praise_2005, berg_slow_2016} or the degrowth movement \cite{engler_15_2024} could prove fruitful. Some platform features, such as the initial design of BeReal to only allow posting once per day and encouraging smaller networks, or the experimental platform "Minus Social" which only allows users to post one hundred posts over their account's lifespan \cite{grosser_minus_2021}, could be seen as steps toward slower, post-growth architectures. Exploring how to resist current extractive, accelerationist, and growth-first mechanisms on social media, either by learning from users about their current practices \cite{zong_data_2024} or by designing novel methods of resistance (such as \cite{escher_hexing_2024}) could also provide important insights.

Building for slower, post-growth, sustainable platforms could help to deal with previously-identified challenges of combating misinformation \cite{wilner_its_2023} and lessen feelings of overwhelm from context collapse \cite{schafer_i_2025}. As the world grapples with ongoing crises, much of which result from unmitigated growth, exploring these redesigns and alternative platform imaginings becomes essential. 

\begin{acks}
Thank you to my coauthors on the works that informed this workshop submission (Dr. Kate Starbird, Annie Denton, Chloe Seelhoff, Jordyn Vo, Lance Garcia, Isha Madan, Alisha Mudbhary, and Ruijingya Tang), and to Dr. Julie Kientz, Dr. Sean Munson, and other members of the UW HCDE departmental seminar who helped me to think through some of these ideas. I am supported by an NSF Graduate Research Fellowship, Grant DGE-2140004. Any opinions, findings, and conclusions or recommendations expressed in this material are those of the author, and do not necessarily reflect the views of the National Science Foundation.
\end{acks}

\bibliographystyle{ACM-Reference-Format}
\bibliography{main}

\end{document}